\documentclass[]{article}
\usepackage{lmodern}
\usepackage{amssymb,amsmath}
\usepackage{ifxetex,ifluatex}
\usepackage{fixltx2e} 
\ifnum 0\ifxetex 1\fi\ifluatex 1\fi=0 
  \usepackage[T1]{fontenc}
  \usepackage[utf8]{inputenc}
\else 
  \ifxetex
    \usepackage{mathspec}
  \else
    \usepackage{fontspec}
  \fi
  \defaultfontfeatures{Ligatures=TeX,Scale=MatchLowercase}
\fi
\IfFileExists{upquote.sty}{\usepackage{upquote}}{}
\IfFileExists{microtype.sty}{%
\usepackage{microtype}
\UseMicrotypeSet[protrusion]{basicmath} 
}{}
\usepackage[margin=1in]{geometry}
\usepackage{hyperref}
\hypersetup{unicode=true,
            pdftitle={Forecasting the Urban Skyline with Extreme Value Theory},
            pdfauthor={Jonathan Auerbach and Phyllis Wan},
            pdfborder={0 0 0},
            breaklinks=true}
\usepackage{graphicx,grffile}
\makeatletter
\def\maxwidth{\ifdim\Gin@nat@width>\linewidth\linewidth\else\Gin@nat@width\fi}
\def\maxheight{\ifdim\Gin@nat@height>\textheight\textheight\else\Gin@nat@height\fi}
\makeatother
\setkeys{Gin}{width=\maxwidth,height=\maxheight,keepaspectratio}
\IfFileExists{parskip.sty}{%
\usepackage{parskip}
}{
\setlength{\parindent}{0pt}
\setlength{\parskip}{6pt plus 2pt minus 1pt}
}
\ifx\paragraph\undefined\else
\let\oldparagraph\paragraph
\renewcommand{\paragraph}[1]{\oldparagraph{#1}\mbox{}}
\fi
\ifx\subparagraph\undefined\else
\let\oldsubparagraph\subparagraph
\renewcommand{\subparagraph}[1]{\oldsubparagraph{#1}\mbox{}}
\fi

\let\rmarkdownfootnote\footnote%
\def\footnote{\protect\rmarkdownfootnote}

\usepackage{titling}

  \title{Forecasting the Urban Skyline with Extreme Value Theory}
  \pretitle{\vspace{\droptitle}\centering\huge}
  \posttitle{\par}
  \author{Jonathan Auerbach and Phyllis Wan}
  \preauthor{\centering\large\emph}
  \postauthor{\par}
  \predate{\centering\large\emph}
  \postdate{\par}
  \date{\today}

\begin{document}
\maketitle

\vfill

\subsection{Abstract}\label{abstract}

The world's urban population is expected to grow fifty percent by the
year 2050 and exceed six billion. The major challenges confronting
cities, such as sustainability, safety, and equality, will depend on the
infrastructure developed to accommodate the increase. Urban planners
have long debated the consequences of vertical expansion---the
concentration of residents by constructing tall buildings---over
horizontal expansion---the dispersal of residents by extending urban
boundaries. Yet relatively little work has predicted the vertical expansion
of cities and quantified the likelihood and therefore urgency of these 
consequences.

We regard tall buildings as random exceedances over a threshold and use
extreme value theory to forecast the skyscrapers that will
dominate the urban skyline in 2050 if present trends continue. We predict
forty-one thousand skyscrapers will surpass 150 meters and 40 floors, an
increase of eight percent a year, far outpacing the expected urban
population growth of two percent a year. The typical tall skyscraper
will not be noticeably taller, and the tallest will likely exceed one
thousand meters but not one mile. If a mile-high skyscraper is
constructed, it will hold fewer occupants than many of the mile-highs 
currently designed. We predict roughly three-quarters the
number of floors of the Mile-High Tower, two-thirds of Next Tokyo's Sky
Mile Tower, and half the floors of Frank Lloyd Wright's The Illinois---three
prominent plans for a mile-high skyscraper. However, the relationship 
between floor and height will vary across cities.

\newpage

\subsection{1. Introduction}\label{introduction}

The world is urbanizing at an astonishing rate. Four billion people live
in urban areas, up from two billion in 1985. By 2050, the United Nations
predicts urban areas will encompass more than six billion residents.
These increases are due to growth in both the world population and the
proportion of the population that resides in urban areas. Roughly half
the world's population is urban, up from forty-percent in 1985 and
projected to rise above two-thirds in 2050 (UN (2018)). The future
preponderance of cities suggests the major challenges confronting
civilization will be urban challenges. Moreover, the particular nature
of these challenges will depend on how cities choose to accommodate 
urbanization (Rose (2016), page 15).

Cities change in response to population growth by either increasing
density---the population per land area---or extending boundaries---the
horizontal distance between city limits. The prevailing paradigm among
urban planners is to preserve city boundaries and encourage density
(Angel et al. (2011)). It argues that density affords certain economies
of scale, such as reducing the cost of infrastructure and social
services like roads, water, safety, and health care. Density is also advocated
to promote sustainability by preserving the city periphery for agriculture or
wildlife (Swilling (2016)). Yet density, if not properly accommodated,
can lead to overcrowding and impede quality of life. Nearly one third of
urban residents in developing regions live in overcrowded slums that
concentrate poverty (UN (2015), page 2).

Some urban planners have argued that density requires vertical expansion,
through the construction of skyscrapers, to prevent overcrowding and
maintain quality of life (Gottmann (1966)), (Al-Kodmany (2012)), (Barr
(2017)). This three-dimensional solution to a two-dimensional problem
was stated as early as 1925 by architect Le Corbusier: ``We must
decongest the centers of our cities by increasing their density''
(Kashef (2008)). In this spirit, economist Glaeser (2011) recommends
policies that ease height restrictions and increase financial incentives
for skyscraper development.

Other urban planners warn that urbanization is too rapid to be
adequately addressed by vertical expansion (James (2001), page 484), 
(Cohen (2006), page 73), (Canepari (2014)). Angel et al. (2011) argue cities
must ``make room'' for urbanization by moving boundaries and
recommend policies that extend the radius of public services like the 
transit system.

It stands to reason that cities will utilize multiple strategies to
accommodate urbanization. For example, cities will incentivize some
vertical expansion and extend the radius of some public services. The
challenges facing cities in 2050 will depend on which policies are
implemented. Anticipating these challenges requires an answer to
questions such as: if present trends continue, how much vertical growth
will the typical city experience by 2050? How far will the typical city
boundary extend?

This paper demonstrates that extreme value theory provides a
principled basis for forecasting vertical growth. It regards tall buildings
as random exceedances over a threshold and uses the statistical laws
governing extreme values to extrapolate the characteristics of
skyscrapers that will dominate the urban skyline in 2050. Similar
arguments have produced successful forecasts in a wide variety of fields,
most notably those concerned with risk management: finance
(e.g. Gencay and Selcuk (2004), Bao et al. (2006), Chan and Gray (2006),
Herrera and Gonz{\'a}lez (2014)) and climate (e.g. Garreaud, RD. (2004),
Ghil (2011), D'Amico et al. (2015), Thompson et al. (2017)). However,
we know of no work that applies these arguments to forecast how cities
will respond to rapid urbanization.

The findings are arranged into five sections. Section 2 describes the 
dataset for this study: a database maintained by the Council on Tall 
Buildings and Urban Habitat (2017) of nearly all skyscrapers worldwide 
taller than 150 meters and completed as of December 2017. A brief 
review of the skyscraper literature follows, motivating the consistency of 
skyscrapers with extreme value theory. Section 3 outlines the log-linear model 
used to conclude that forty-one thousand skyscrapers will be completed 
by 2050. Section 4 outlines the generalized Pareto distribution (GPD)
used to conclude that there is a seventy-five percent chance a 
skyscraper will exceed one thousand meters by 2050 and a nine percent
chance it will exceed a mile. Section 5 outlines the censored asymmetric
bivariate logistic distribution used to conclude that a mile-high skyscraper, 
if built, will have around 250 floors. The paper concludes in the final two 
sections by discussing some methodological and policy consequences 
of these predictions and identifying areas for future research.

\subsection{2. Extreme Value Theory is Consistent with Skyscraper
Theory}\label{extreme-value-theory-is-consistent-with-skyscraper-theory}

This paper predicts the prevalence and nature of skyscrapers in the year
2050 if present trends continue. The present section reviews the
definition of a skyscraper and motivates the use of extreme value
theory to characterize tall buildings.

Strictly speaking, the term skyscraper refers not to height but to the
mode of construction. A skyscraper is defined as any multi-story
building supported by a steel or concrete frame instead of traditional
load-bearing walls (Curl and Wilson (2015), page 710). The tall
buildings capable of sustaining the dense cities of the future will
almost certainly be skyscrapers. Over the past century, building beyond
a few floors has required a supporting frame to be economical.

In contrast to the precise definition of a skyscraper, the definition of
a tall building depends on context. From a safety perspective,
high-rises---multi-story buildings beyond 23 meters (75 feet)---are harder to evacuate
than low-rises. For characterizing the urban skyline, however, a twenty story skyscraper
in Sydney, Australia might go unnoticed while the same building in Sidney,
New York would dominate the skyline. The Council on Tall Buildings and Urban
Habitat (CTBUH) sets international standards for the purposes of research and
arbitrating titles, such as the world's tallest building. They designate buildings
higher than fifty meters as tall, three-hundred meters as supertall, and six-hundred
meters as megatall. Height is defined as the distance from ``the level of the lowest,
significant, open-air, pedestrian entrance to the architectural top of the
building, including spires, but not including antennae, signage, flag poles
or other functional-technical equipment. This measurement is the most
widely utilized \ldots{}'' (Council on Tall Buildings and Urban Habitat
(2017)). The New York City Fire Department relies on all of these definitions
when inspecting buildings, dispatching firefighters, and conducting investigations.

CTBUH also maintains The Skyscraper Center, a database of every tall
building worldwide. The data is recognized as ``the premier source for
accurate, reliable information on tall buildings around the world'' (ibid.)
However, the data relies on CTBUH members and the public to add entries,
resulting in the partial or complete omission of some smaller buildings.
For this reason, this paper refers to buildings exceeding 150 meters and
40 floors as tall, roughly the size of the United Nations Headquarters
in New York City. CTBUH catalogs 3,251 tall skyscrapers in 258 cities as
of December 2017. In later sections, it will also be convenient to 
consider skyscrapers exceeding 225 meters or 59 floors, roughly the size 
of One Penn Plaza in New York City. Due to the nature of our analysis, we 
refer to these building as extremely tall. CTBUH catalogs 325 extremely
tall skyscrapers in 81 cities as of December 2017.

The height and year each skyscraper was completed is displayed in the
top-left panel of Figure \ref{fig:numsky}. From the panel, it appears as
though a simple statistical relationship might govern skyscraper
heights, ignoring the hiatus between the Great Depression and the Second
World War, which is assumed to be anomalous. But it cannot be taken for
granted that skyscrapers can be modeled statistically. These buildings
are modern marvels, requiring enormous cooperation across teams of
architects, engineers, financers, and multiple levels of government.
Extrapolation from the data is only meaningful if the determinants
underlying skyscraper construction are varied and independent enough to
be characterized statistically. The remainder of this section outlines
an argument supporting this view.

Skyscrapers appeared in the nineteenth century only after technological
innovations, such as the elevator brake and the mass production of
steel, made building beyond a few floors economical. By the turn of the
twentieth century, a handful of twenty floor buildings had been
completed in major cities across the United States. By the mid-twentieth
century, engineering advances allowed for mile-high buildings (1609.34
meters), and various architects have proposed designs, such as the
Houston Pinnacle (1 mile, 500 floors), the Ultima Tower (2 miles, 500
floors), and the Sky Mile Tower (1 mile, 400 floors). Perhaps most
famously, architect Frank Lolyd Wright proposed a mile-high skyscraper, The Illinois, 
in 1957, planning for 528 floors (Council on Tall Buildings and Urban
Habitat (2017)).

But none of these designs have been realized. Despite the technical
ability to build tall, mile-high designs are considered impractical
because they are unlikely to turn a profit. Indeed, highly ambitious
projects often fail for financial reasons. (Lepik (2004), pages 21-2) For 
example, the Jeddah Tower was originally planned for one mile 
(330 floors). After the Great Recession, the height was reduced by more
than a third to one thousand meters (167 floors).

Skyscrapers are a commercial response to an economic phenomenon. They
arise when land values produce rents that exceed the enormous cost of
construction and maintenance. Aesthetics are a secondary concern (Clark
and Kingston (1930)), (Willis (1995)), (Ascher and Vroman (2011), pages
12, 22). Indeed, it was in the aftermath of the 1871 Chicago fire
that the urgent need for new office space produced the First Chicago
School of skyscrapers (Lepik (2004), page 6). Skyscraper construction
has since followed the rise of oil-rich Middle Eastern countries in the
1980s, the former Soviet-bloc countries in the 1990s, and the Pacific
Rim countries in the 2000s (Sennott (2004) , page 1217). Barr and Luo
(2017) find that half of the variation in China's skyscraper
construction can be explained by population and gross city product
alone.

The enormous cost of a skyscraper is not only the result of additional
construction materials. Taller buildings require concrete to be pumped
at higher pressures (Ascher and Vroman (2011), page 83). Nor is the cost
entirely in raising the building itself. Taller buildings require more
elevators, which reduces the floor area available for occupancy and thus
the revenue potential of the building (ibid., page 33). Human comfort is also 
a factor. Excessive elevator speed (ibid., page 103) and building sway (ibid., 
page 61) can produce motion sickness even if safe. Additional
considerations include government policy (permits and zoning, financial
incentives, and public infrastructure), culture (aesthetic, equity, and
sustainability), and environment (foundation quality, prevailing winds,
and natural disaster frequency).

In short, a litany of factors must align favorably to produce a
skyscraper. One could hypothetically account for every factor in every
city property and predict the future of skyscraper development. Yet it
is convenient, and perhaps more accurate, to regard unobserved
heterogeneity among these factors as a source of randomness that obeys
the laws of probability. Statistical models can then be used to quantify
the probability a property possesses the factors that will produce a
particular skyscraper where the parameters of the models are determined
from data.

Extreme value theory provides a principled strategy for choosing a
statistical model. The theory is simplified by assuming skyscraper
development is independent. This assumption is certainly violated for
contemporaneously completed skyscrapers within a city since many of the
aforementioned factors are correlated within a city at a specific period
in time. But a series of investigations by economist and skyscraper
expert Barr suggest that the conditions underlying skyscraper
development are largely independent across cities and time periods. For
example, Barr (2012) finds competition within cities is limited to
periods close in time and space. Barr, Mizrach, and Mundra (2015) find
skyscraper height is not a useful indicator of economic bubbles or
turning points. Barr and Luo (2017) find little evidence that cities in
China compete for the tallest building.

Considerable economic pressure produces the factors that drive
skyscraper development, and the catalyst for this pressure has varied
idiosyncratically by city and time period. It is perhaps because of this
idiosyncratic variation---and the fact that no city possesses more than ten
percent of all skyscrapers---that the prevalence and nature of
skyscrapers so closely follows the distributions predicted by extreme
value theory, as demonstrated in the following sections, despite periods
of economic and political turmoil within nearly every city since the
Second World War.

\subsection{3. Predicting the Quantity of Skyscrapers Completed by
2050}\label{predicting-the-quantity-of-skyscrapers-completed-by-2050}

Skyscraper construction has increased at a remarkably steady rate. The
number of tall skyscrapers---skyscrapers exceeding 150 meters and 40
floors---has risen eight percent each year since 1950. If this trend continues,
the eight percent annual growth rate of skyscrapers will far outpace the two
percent annual growth rate of urban populations. Forty-one thousand tall
skyscrapers will be completed by 2050, and if the UN's population
predictions hold, there will be roughly 6,800 per billion city residents
in 2050 compared to the roughly 800 per billion city residents today.

The eight percent rate was determined by fitting the following
log-linear model. Let \(N_t\) be the number of skyscrapers completed in
year \(t\), \(t = 1950,\ \ldots,\ 2017\). The \(N_t\)'s are assumed to be
independent and follow a Poisson distribution with mean

\[E[N_t] = \exp(\alpha + \beta t)\]

The Poisson distribution can be justified theoretically by regarding
skyscraper development as the sum of independent Bernoulli trials.
Suppose \(D(t)\) is a dataset containing the height of every building in
the world completed in year \(t\). Define \(D_u(t)\) to be the subset of
all buildings in \(D(t)\) exceeding the height threshold \(u\). For
sufficiently large \(u\), the number of skyscrapers in \(D_u(t)\),
\(N_t\), is well approximated by a Poisson distribution. See Coles et
al. (2001) (page 124) for a more detailed discussion of the Poisson
Process limit for extremes.

\begin{figure}
\centering
\includegraphics{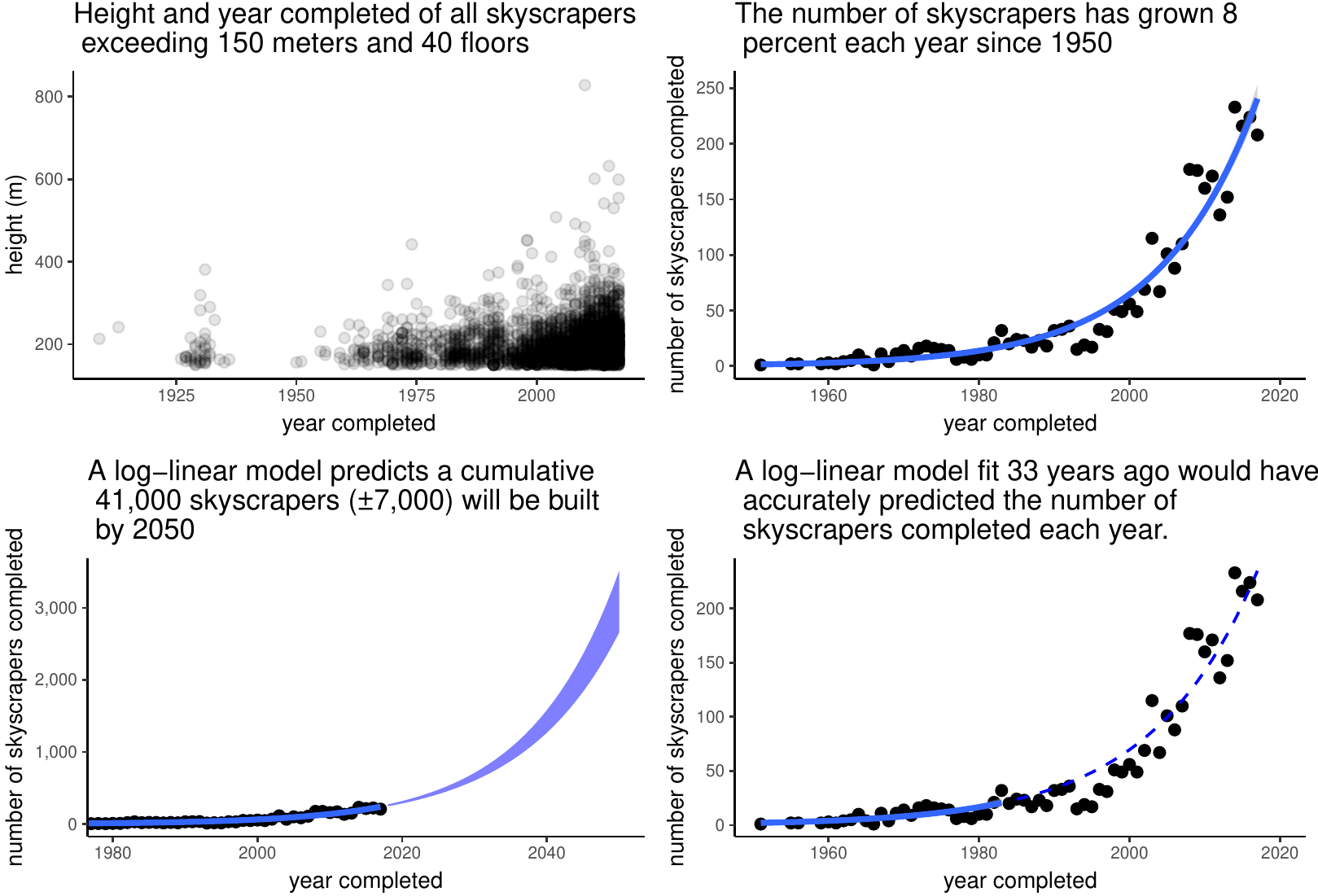}
\caption{\label{fig:numsky}We estimate the quantity of skyscrapers in
2050, thirty-three years after the data was collected at the end of
2017. The top-left panel shows the height (meters) and year all 3,251
skyscrapers exceeding 150 meters and 40 floors were completed. The
top-right panel shows the number of skyscrapers completed each year
(points) and the expected number after fitting a log-linear model (blue
line). The bottom-left panel extrapolates the number of new skyscrapers
to be completed each year until 2050. The bottom-right panel shows that,
had the same prediction been performed thirty-three years ago at the end
of 1984, an estimated 3,082 new skyscrapers would have been completed by
2018. This prediction would have been three percent more than the actual
number.}
\end{figure}

We use the \texttt{glm} function in the R Core Team (2018) package 
\texttt{stats} to choose parameter estimates \(\hat \alpha\) and
\(\hat \beta\) that maximize the likelihood and to calculate their
standard errors. The plug-in estimate,
\(\hat E[N_t] = \exp(\hat \alpha + \hat \beta t)\), is plotted against
the data in the top-right panel of Figure \ref{fig:numsky}. An inner 95
percent predictive distribution for the years 2020 to 2050 is added in
the bottom-left panel. A cumulative thirty-eight thousand skyscrapers is
estimated for completion between 2018 and 2050 if present trends
continue (forty-one thousand total, with a standard error of seven
thousand), about twelve times the current number.

The top-right panel of Figure \ref{fig:numsky} exhibits short-term
serial correlation in the residuals. Most noticeably, the historic trend
was not maintained in the late 1990s. Nevertheless, the log-linear
relationship remains predictive. To demonstrate the accuracy of the
model for predicting 2050, thirty-three years after the data was
collected in 2017, we predicted the last thirty-three years using data
up until 1984. The bottom-right panel shows that if such predictions had
been made in 1984 (dotted blue line), they would align closely with the
actual number of skyscrapers built each year. The log linear model
anticipates 3,082 skyscrapers by the end of 2017 when in fact 2,988 were
built between 1950 and 2017, a difference of three percent.

\subsection{4. Predicting the Height of Skyscrapers Completed by
2050}\label{predicting-the-height-of-skyscrapers-completed-by-2050}

The tallest skyscraper has doubled in height since 1950, yet the height
increase of the typical extremely tall skyscraper is not statistically
significant. The same distribution describes the height of skyscrapers
exceeding 225 meters since 1950. We conclude the tallest skyscraper
is increasing not because skyscrapers are getting taller but because
more buildings are being constructed and thus more buildings are eligible
to be the tallest. Assuming this distribution continues to describe the height of
skyscrapers completed by 2050, the probability a new building will exceed the current
tallest building---the Burj Khalifa (828 meters)---is estimated to be nearly
100 percent. The probability that a new building will exceed the Jeddah
Tower (1,000 meters)---scheduled to open in 2020---is 77 percent. The
probability that a new building will exceed one mile is 9 percent.

The distribution of skyscraper heights was approximated by the family of
generalized Pareto distributions (GPD). These distributions can be
justified both theoretically---by regarding tall buildings as random
exceedances over a threshold---and numerically---by comparing the
observed skyscraper heights to random draws from a GPD. We begin with
the theoretical justification. Let \(X\) be a random variable with
unknown distribution function \(F\). The exceedance conditional distribution 
describes the behavior of \(X\) given that it exceeds a large threshold

\[F_u(y) = \Pr(X-u\le y|X>u)\]

for \(u\) large. The Pickands-Balkema-de Haan Theorem (Balkema and De
Haan (1974), Pickands III (1975)) states that for a sufficiently
large class of distributions \(F\),
\[F_u(y) \to H(y) = 1 - \left(1+ \frac{\xi y}{\sigma}\right)^{-1/\xi},\quad \mbox{as $u\to\infty$},\]

where \(\sigma>0\), \(y>0\) when \(\xi>0\) and
\(0\le y\le -\frac{\sigma}{\xi}\) when \(\xi<0\). For \(\xi=0\), the
value of the function is taken as its limit \(H(y)=1-\exp(-y/\sigma)\).
The set of distribution functions \(H(\cdot)\) is the GPD family. See 
Coles et al. (2001) (page 74) for a more detailed discussion of the 
GPD for modeling threshold exceedances.

Suppose \(D\) is a dataset of observations, and \(D_u \subset D\) is the
subset of all observations whose measurement, X, is greater than threshold 
\(u \gg 0\). The Pickands-Balkema-de Haan Theorem justifies approximating the
distribution of each $X$ with GPD
\[\Pr(X\le x|X>u) = 1 - \left(1+ \frac{\xi (x-\mu)}{\sigma}\right)^{-1/\xi},\]
where the parameters \((\mu,\xi,\sigma)\) can be estimated from dataset
\(D_u\) by maximizing the likelihood.

As in the previous section, \(D\) is taken to be the dataset containing
the height of every building in the world, and \(D_u\) is the subset of
all buildings in \(D\) exceeding the height threshold \(u\). However,
unlike the previous section, the default threshold of \(u = 150\) meters
may not be sufficiently high to reliably extrapolate the tail of \(D\).
We note that, were \(u\) sufficiently high so that \(X-u|X>u\) followed
a GPD, \(X-u'|X>u'\) would also follow a GPD with the same \((\sigma,\xi)\)
for any \(u'>u\). This suggests the following strategy for choosing
\(u\). Produce a sequence of candidate thresholds \(u_1<\ldots< u_d\)
and estimate the GPD parameters \((\mu_{u_i},\sigma_{u_i},\xi_{u_i})\),
yielding \((\hat\mu_{u_i},\hat\sigma_{u_i},\hat\xi_{u_i})\). Then select any
threshold \(u\) for which the estimates \(\hat\xi_{u'}\) and
\(\hat\sigma_{u'}\) remain stable for all \(u'>u\). See Gencay and Selcuk
(2004) for a general discussion of GPD threshold choice and prediction.

\begin{figure}
\centering
\includegraphics{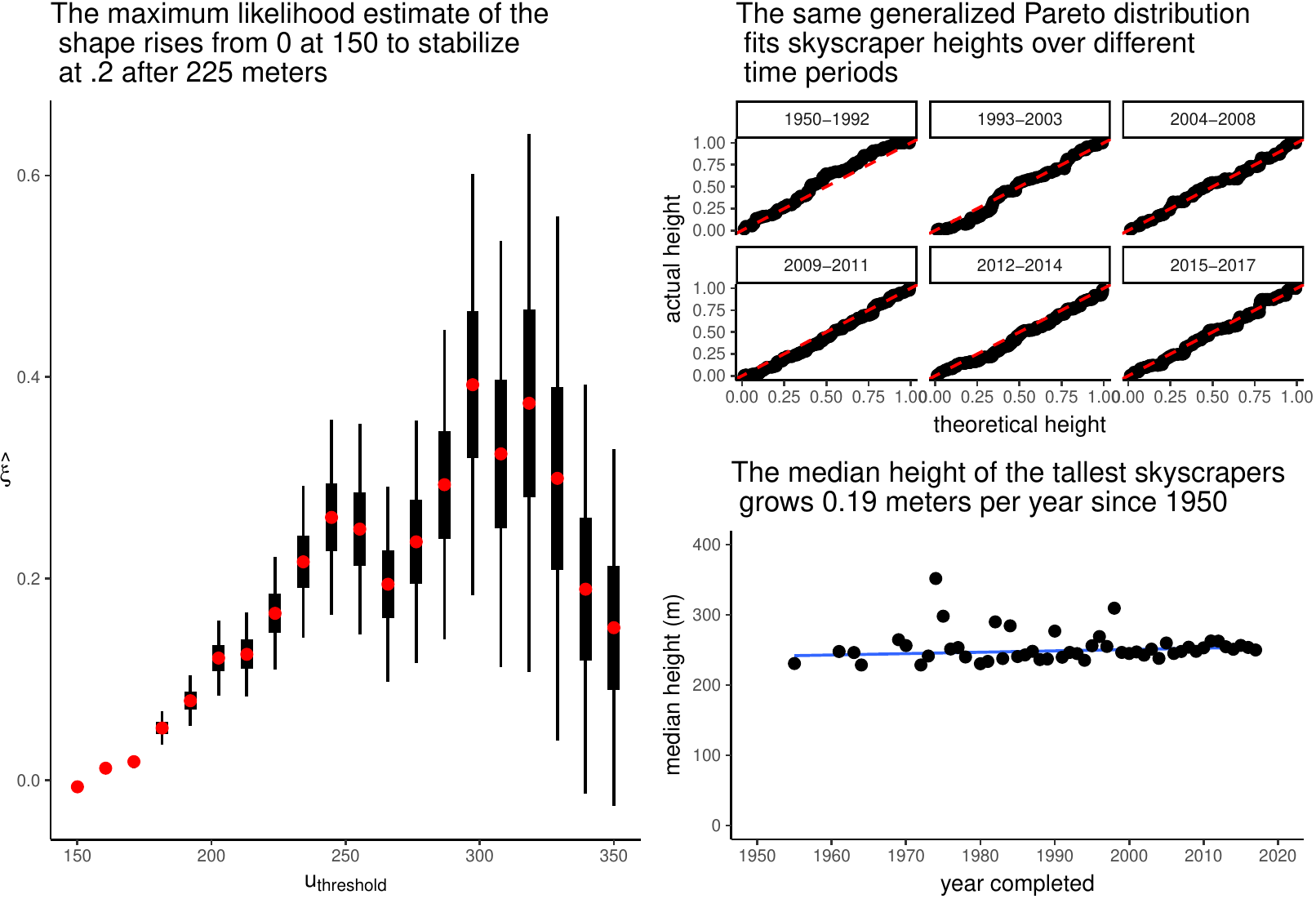}
\caption{\label{fig:qq}We estimate the height of extremely tall skyscrapers
in 2050. The left panel shows estimates and confidence intervals (50 and
95 percent) for the maximum likelihood estimate of the GPD shape
parameter, excluding observations below different thresholds. The
estimate stabilizes at around .2 for thresholds above 225 meters. The
GPD estimates using the 225 threshold are retained for the top-right
plot, which shows q-q plots of transformed skyscraper heights against
simulations from a standard uniform distribution for six time periods
(sextiles). The transform is the cumulative distribution of the
generalized Pareto distribution. These plots indicate the same
generalized Pareto distribution accurately describes the distribution of
skyscraper heights exceeding 225 meters at different time periods. The
bottom right-panel shows the median skyscraper height above the 225
meter threshold has not increased significantly over the last forty
years. Median regression (blue line) indicates extremely tall
skyscrapers grow .19 meters each year with a p-value of .41. We conclude
the typical extremely tall skyscraper will not increase noticeably by 2050.}
\end{figure}

The top-left panel of Figure \ref{fig:qq} shows maximum likelihood
estimates for the shape parameter, \(\hat \xi_{u}\), when fit to skyscraper
heights exceeding a sequence of thresholds from 150 to 350
meters. Point estimates are colored red, and 50 (95) percent confidence
intervals are depicted with thick (thin) lines. The point estimates
increase as the threshold increases from 0 at 150 meters and stabilize
around .2 after 225 meters. Hill plots (not shown) also indicate a
stable \(\hat \xi\) of .2. In contrast, the estimate for the scale
parameter, \(\hat \sigma\), changes little as the threshold is
increased. Since the extreme value assumptions appear satisfied at the
225 meter threshold, we refer to skyscrapers exceeding this height as
``extremely tall''. Computation is discussed further in the Appendix.

A numeric justification for the GPD can be made by comparing simulations
of skyscraper heights to the observed data. In fact, at the threshold of 225, 
simulations from the same GPD describe the heights of skyscrapers completed at 
different time periods. To demonstrate this, we first obtain the maximum likelihood 
estimates for the GPD parameters, \(\hat \mu_{225}\), \(\hat \sigma_{225}\), and
\(\hat \xi_{225}\). We then partition the skyscrapers built after 1950 into
sextiles according to the year in which they were completed. The
top-right panel of Figure \ref{fig:qq} shows q-q plots for the
skyscrapers in each time period. However, instead of plotting skyscraper
heights against the heights predicted from a GPD with parameters
\(\hat \mu_{225}\), \(\hat \sigma_{225}\), and \(\hat \xi_{225}\), we
transform both axes by the GPD cdf,
\(1-[\hat \xi_{225}(X - \hat \mu_{225})/\hat \sigma_{225} + 1)]^{(-1/\hat \xi_{225})}\).
After transformation, the theoretical distribution is a standard
uniform. We find these q-q plots more stable than the unadjusted plot, and it is
easier to assess the fit.

The close fit in each time period suggests the distribution of
extremely tall skyscraper heights does not change over time. This is 
investigated further in the bottom-right panel of Figure \ref{fig:qq}, which shows
the median height of extremely tall skyscrapers each year since 1950. A median
regression line is estimated using the R Core Team (2018) package \texttt{quantreg}
(Koenker (2018)). The height of the typical extremely tall
skyscraper increases less than half a meter each year, a relatively
small 3.6 percent increase over sixty-eight years. The standard errors of the 
regression parameters are also retained, and the increase is not statistically
significant (p-value .41).

\begin{figure}
\centering
\includegraphics{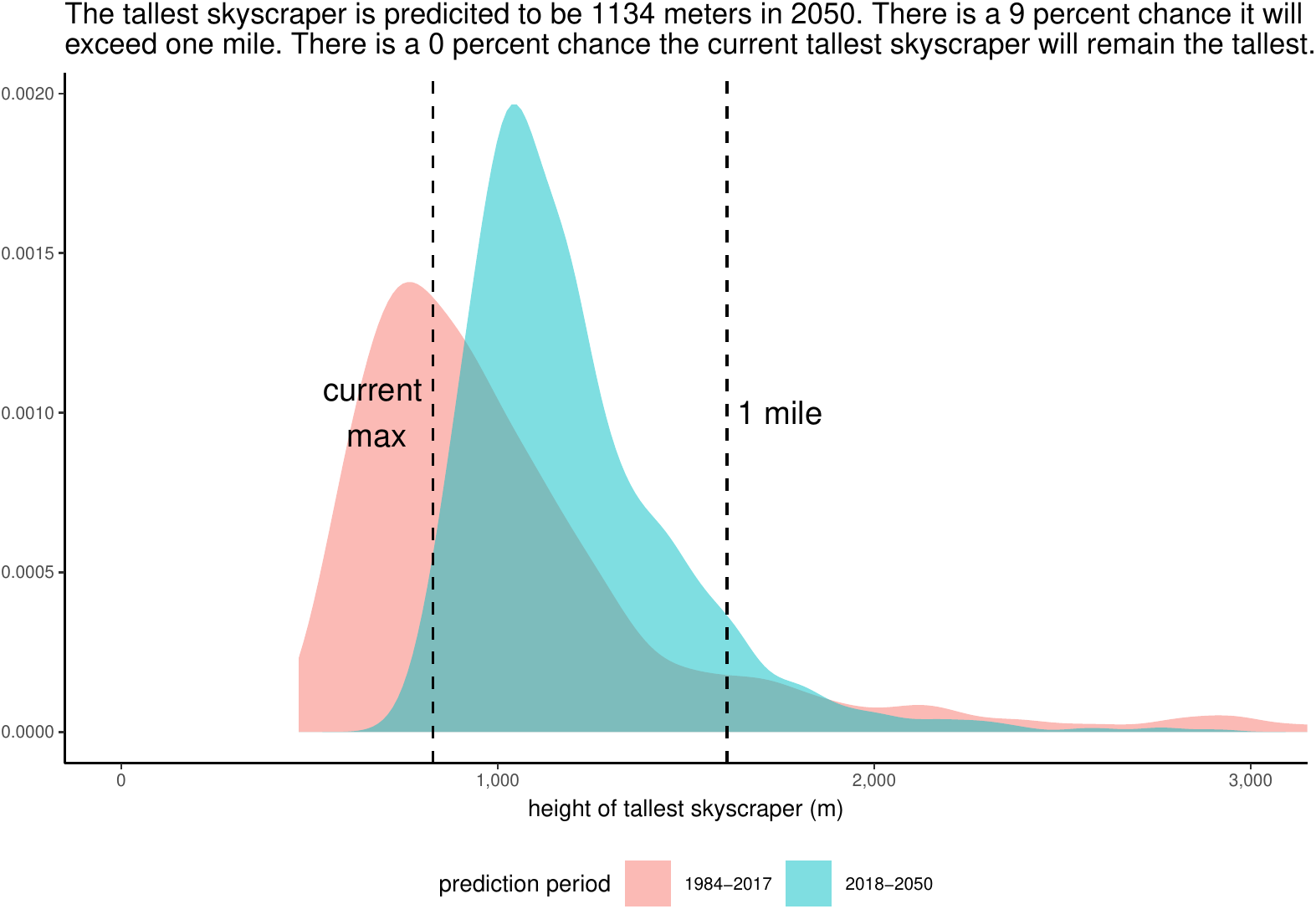}
\caption{\label{fig:max_height}We simulate the height of the
tallest skyscraper in 2050. The simulated density (red) suggests the tallest
building in the world is unlikely to exceed one mile (dashed line on
right side). However, it will almost certainly be taller than the
current tallest building, the Burj Khalifa (828 meters, dashed line on
left side) and likely taller than the Jeddah Tower (one thousand
meters), expected for completion in 2020. Had the same simulation been
conduced in 1984, the density (blue) would have found the tallest skyscraper in 2018 to
be between six hundred and one thousand meters.}
\end{figure}

We conclude that the urban skyline is driven primarily by the
exponential increase in the number of buildings completed each year.
Years with more construction are more likely to build extremely tall 
skyscrapers, and the increase of the typical skyscraper is negligible by comparison.
For the purpose of predicting the height of the tallest skyscraper in 2050, we 
assume this trend will continue. We use the \texttt{gpdSim} function in the R 
Core Team (2018) package \texttt{fExtremes} (Wuertz, Setz, and Chalabi (2017)) 
to draw from the GPD with parameters \(\hat \mu_{225}\), \(\hat \sigma_{225}\), and
\(\hat \xi_{225}\), approximately 8,400 times, where 8,400 is the number
of new skyscrapers estimated to be completed by 2050 in the previous
section multiplied by the empirical probability a skyscraper above 150
meters will exceed 225 meters. The maximum height is retained, and this
process is repeated one thousand times. Figure \ref{fig:max_height}
shows the resulting distribution of maximum heights (blue). A right-sided 95 percent
predictive interval ends at 1,800 meters.

To demonstrate the accuracy of this approach for predicting 2050, thirty-three 
years after the data was collected in 2017, we conduct a second simulation using only 
data that would have been available before 1984. The 
GPD is fit using all skyscrapers above the 225 threshold, and we repeatedly 
simulate the maximum skyscraper one thousand times. These simulations assume
approximately three thousand tall skyscrapers will be built between 1984
and 2017, around fifteen percent of which will exceed 225 meters. The
distribution of the maximum height predicted in 1984 is shown in
Figure \ref{fig:max_height} (red). We find that the current tallest
skyscraper, at 828 meters, would have been considered likely. The median
simulation is 924 meters, roughly ten percent above this value.

\subsection{5. Predicting the Number of Floors in Skyscrapers Completed
by
2050}\label{predicting-the-number-of-floors-in-skyscrapers-completed-by-2050}

The marginal number of floors in the typical skyscraper decreases as
height increases. Height alone overstates the ability of skyscrapers to 
accommodate a growing population. Assuming the marginal
number of floors continues to decrease as skyscraper height increases, the 
one thousand meter building is estimated to have seventy
percent the floors of the mile-high building---despite being sixty-two
percent of the height. While diminishing marginal floors is reflected in
most architectural designs, we find the estimated number of floors will
diminish faster with height than most designs anticipate. However, the
exact relationship between height and number of floors will vary by city.

The relationship between height and number of floors is extrapolated to
extremely tall skyscrapers using the following bivariate extreme value
model. Let \((X,Y)\) be a bivariate random variable with GPD margins 
as justified in Section 4. Denote the respective parameter sets indexing
the GPDs as \((\mu_x,\sigma_x,\xi_x)\) and \((\mu_y,\sigma_y,\xi_y)\), and 
apply the monotone transformations

\begin{eqnarray} 
\tilde{X} &=& - \log \left( \left[1+\frac{\xi_x(X-\mu_x)}{\sigma_x}\right]^{-1/\xi_x}\right) \label{eq:transform:x}\\
\tilde{Y} &=& - \log \left( \left[1+\frac{\xi_y(Y-\mu_y)}{\sigma_y}\right]^{-1/\xi_y}\right) \label{eq:transform:y}
\end{eqnarray}

such that \((\tilde{X},\tilde{Y})\) has standard exponential margins. The joint distribution \((\tilde{X},\tilde{Y})\) is 
modeled using the asymmetric logistic distribution, where given thresholds \(u, v \gg 0\), for \(X > u, Y > v\),

\begin{equation} \label{eq:alog}
\Pr(\tilde{X}>{x}, \tilde{Y}>{y}) \,\propto\, \exp{\left(-(1-\theta_x){x} - (1-\theta_y){y} - ({x}^r\theta_x^r +{y}^r\theta_y^r)^{1/r}\right)}, \quad \theta_x,\theta_y\in[0,1], \quad r\ge 1.
\end{equation}

This model has the advantage of being simple and flexible, and it is
suitable for larger sample problems, as in our case. Tawn (1988)
and Coles et al.~(2001) (page 142) provide a detailed discussion of 
bivariate models for threshold exceedances.

The thresholds \(u = 225\) and \(v = 59\) are selected based on the
stability of the marginal distribution as described in Section 4. The nine 
parameters $\mu_x$, $\sigma_x$, $\xi_x$, $\mu_y$, $\sigma_y$, $\xi_y$, $\theta_x$, $\theta_y,$ and $r$ are estimated 
using the heights and floors of all skyscraper exceeding 225 meters or 59 floors, maximizing 
the censored likelihood:
\begin{eqnarray*}
L_c(\mu_x,\sigma_x,\xi_x, \mu_y,\sigma_y,\xi_y, \theta_x,\theta_y,r)&=& \prod_{x_i>u,y_i>v} f(x_i,y_i)P(X>u,Y>v)  \\
&& \quad\prod_{x_i\le u,y_i>v} f_{Y}(y_i)P(x\le u) \prod_{x_i>u,y_i\le v}  f_{X}(x_i)P(y\le v),
\end{eqnarray*}
where $f,f_X,f_Y$ are the joint and marginal density function of $(X,Y)$ from the 
transformations \eqref{eq:transform:x} and \eqref{eq:transform:y} and the distribution
function \eqref{eq:alog}.

\begin{figure}
\centering
\includegraphics{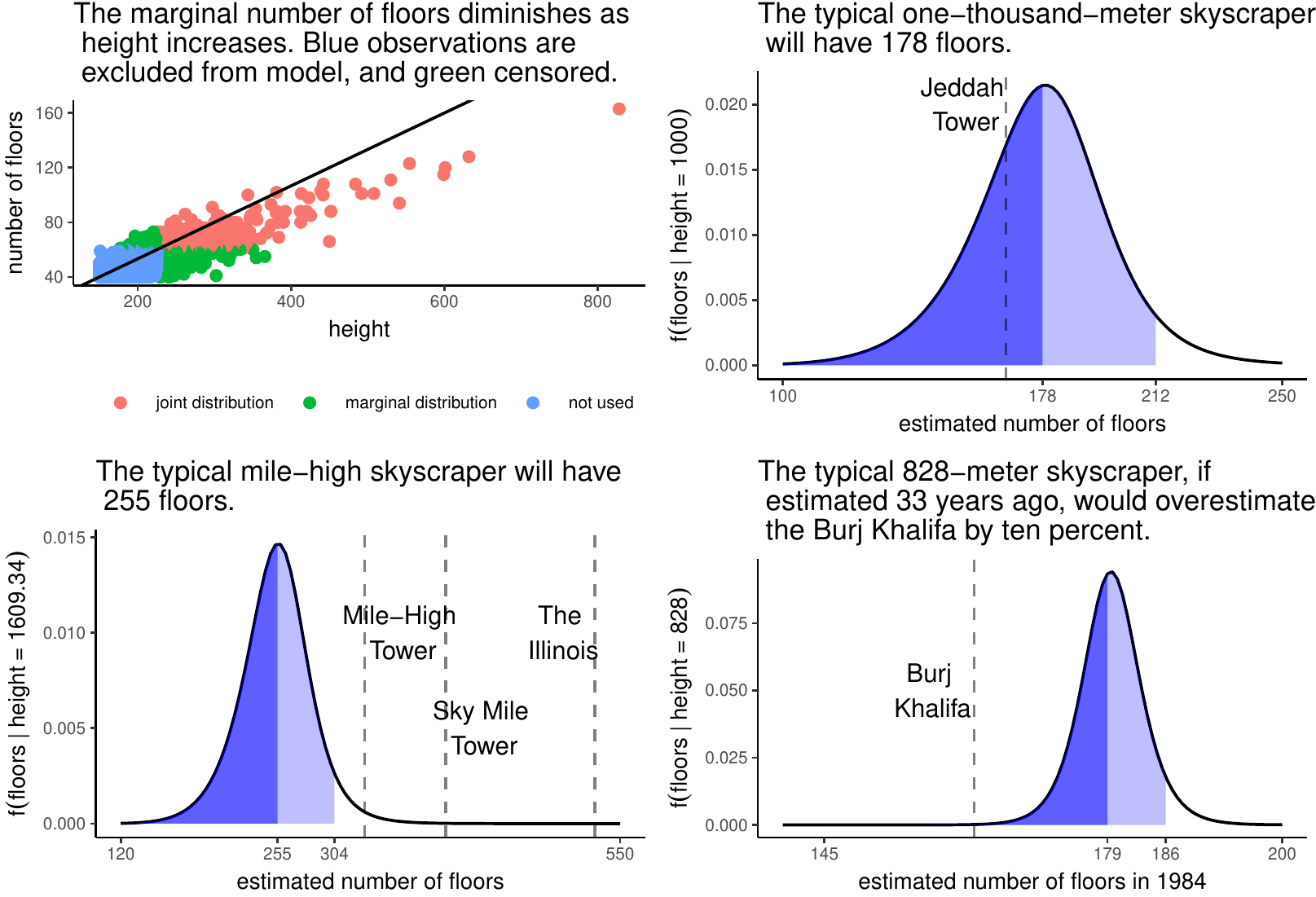}
\caption{\label{fig:max_flrs}We estimate the number of floors in the
tallest skyscrapers in 2050. The top-left panel shows the height
(meters) and number of floors for each skyscraper exceeding 150 meters
and 40 floors. The black line represents the floor to height ratio of
the median skyscraper. This ratio is not preserved as height increases.
A bivariate asymmetric logistic distribution is used to approximate the
joint distribution, and colors depict the use of each observation for
estimating the parameters. Skyscrapers below 225 meters and 59 floors
(blue) are not used to estimate the parameters. Skyscrapers above 225
meters and 59 floors (red) are modeled using the complete likelihood.
The remaining skyscrapers (green) are modeled using the censored
likelihood. The joint distribution indexed by the maximum likelihood
parameters is then used to make probabilistic statements of future
skyscrapers. The bottom-left (top-right) panel shows the conditional
density of the number of floors for a one-mile (one-thousand-meter) tall
skyscraper. Right-side 50 (95) percent intervals are shaded dark (light)
blue. This distribution is compared with actual skyscraper plans (dotted
line). For example, the one-mile skyscraper will have roughly
three-quarters the number of floors in the Mile-High Tower, two-thirds
in Next Tokyo's Sky Mile Tower, and half the floors in Frank Lloyd
Wright's The Illinois. The bottom-right panel shows that, had the same
analysis been conducted in 1984, it would have estimated the Burj
Khalifa to have 179 floors, ten percent more than it actually had when
completed twenty-four years later.}
\end{figure}

The top left panel of Figure \ref{fig:max_flrs} displays the height and number of
floors of every tall skyscraper, colored by its contribution to the censored
likelihood, $L_c$. Skyscrapers below 225 meters and 59 floors (blue) do not contribute to the likelihood
and are not used to estimate the parameters. Skyscrapers exceeding 225 meters and 59 floors (red)
make up the first factor. The remaining skyscrapers (green) make up the second two factors. For
example, a 250 meter skyscraper with 50 floors is treated like a 250 meter skyscraper whose
floors are only known to be below 59. This approach is similar to the censored likelihood in
Huser et al.~(2016), except that skyscrapers at or below 225 meters and 59 floors are excluded from
the analysis. Computation is discussed further in the Appendix.  

The maximum likelihood parameters are retained to estimate the
conditional density of the number of floors for a one-thousand-meter
skyscraper (Figure \ref{fig:max_flrs}, top-right panel) and a one-mile tall
skyscraper (Figure \ref{fig:max_flrs}, bottom-left panel). Dark blue
regions represent the right-sided 50 percent region, and light
blue regions represent the right-sided 95 percent region. These
densities are compared with actual skyscraper plans (dotted line). The
median one-thousand-meter skyscraper is estimated to have 107 percent
the number of floors of the Jeddah Tower (to be completed in 2020). The
median one-mile skyscraper is estimated to have roughly three-quarters
the number of floors of the Mile-High Tower, two-thirds of Next Tokyo's
Sky Mile Tower, and half the floors of Frank Lloyd Wright's The
Illinois.

As in the previous two sections, the same analysis is performed with
only data available in 1984. The bottom-right panel shows the
conditional density of the 828 meter skyscraper as it would have been
estimated using the threshold of 225 meters or 59 floors. In 1984, an 828
meter skyscraper would have been nearly twice the height of the current
tallest building, the Willis Tower (then Sears Tower, 442 meters and 108 floors). The
conditional median predicts the typical 828 meter skyscraper would have
179 floors, ten percent more than the Burj Khalifa completed
twenty-four years later. Simply scaling the Willis Tower to the height of
the Burj Khalifa would yield 202 floors, overestimating the actual number by
twenty-four percent. A linear regression model fit with extremely tall skyscrapers 
overestimates by fifteen percent.

\begin{figure}
\centering
\includegraphics{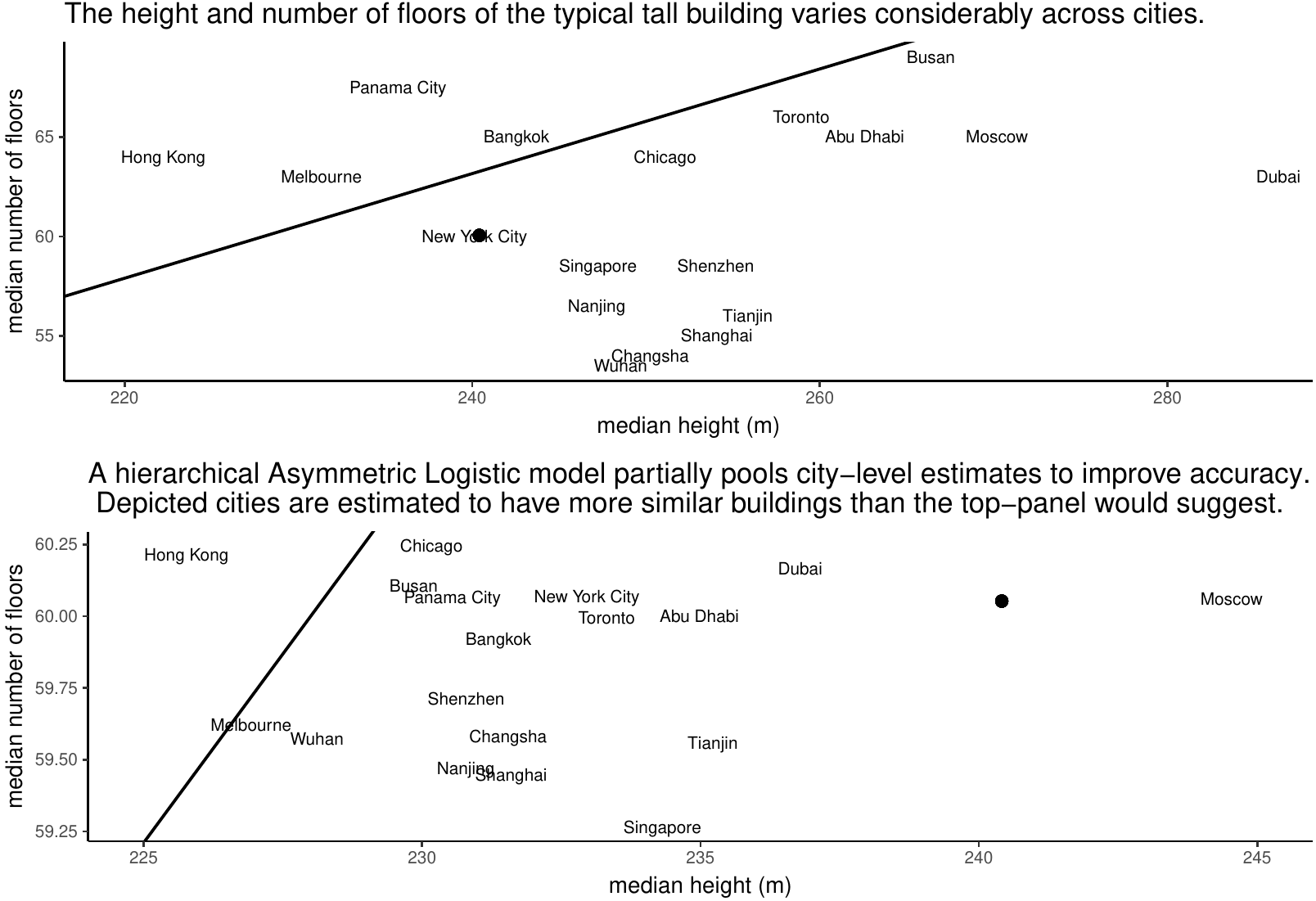}
\caption{\label{fig:random_alog}We estimate the height and number of
floors of the tallest buildings of major cities in 2050. The top panel
shows the empirical median height (m) and number of floors for extremely
tall skyscrapers (exceeding 225 meters or 59 floors) in select cities. 
The medians are based on roughly ten observations each and are likely
unreliable estimates of skyscrapers in 2050. More accurate city-level
predictions can be made using a hierarchical bivariate asymmetric
logistic distribution, where we allow the marginal GPD parameters to
vary by city according to a normal distribution with unknown mean and
variance. All parameters are then estimated by maximizing the
likelihood. The bottom panel shows the estimated median for the select
cities in the hierarchical model. The median of the non-hierarchical
model in the previous Figure is represented by a black dot. These
city-level estimates can be seen as a compromise between the noisy
empirical medians in the top panel and the accurate, but global median
estimated by the non-hierarchical model. The line reflects the 3.8 meter
per floor of the typical tall skyscraper.}
\end{figure}

The estimated relationship between floor and heights varies considerably
across cities. The top panel of Figure \ref{fig:random_alog} shows the
empirical median height (meters) and number of floors of extremely tall
skyscrapers (exceeding 225 meters or 59 floors) in select cities. The
medians are each based on roughly ten observations, and thus sampling
variation overstates the likely differences between the typical extremely tall 
skyscrapers of cities in the year 2050.

We augment the bivariate model to estimate city-level
medians. We allow the marginal GPD parameters to vary by city,
according to a normal distribution with an unknown mean and variance.
Such hierarchical models are often used to produce city-level estimates
that have smaller errors on average than the corresponding stratified
estimates. The use of a parameter hierarchy also has a Bayesian
interpretation. See Coles et al. (2001) (page 169) and Vehtari (2017)
for two discussions of Bayesian inference and extremes.

The bottom panel of Figure \ref{fig:random_alog} shows the estimated
median for the select cities in the hierarchical model with parameters
selected by maximum likelihood. The median of the non-hierarchical model
in the previous Figure is represented by a black dot. (Note that Hong Kong
and New York City contain a disproportionately large number of
extremely tall skyscrapers.) These city-level estimates can be seen as a
compromise between the noisy empirical medians in the top panel and the more
accurate but global median estimated by the non-hierarchical model. However,
despite partial pooling across cities, the typical height per floor ratio still spans a
considerably large range: 3.6 (Hong Kong) to 4 (Moscow).

\subsection{6. Discussion}\label{discussion}

This paper applied extreme value analysis to predict the prevalence 
and nature of skyscrapers if present trends continue until the year 2050. 
The findings of these sections have both methodological and policy 
consequences. This section discusses the methodological consequences, 
while the following section considers policy.

Section 4 found that skyscraper heights are well described by a generalized
Pareto distribution with a positive shape, \(\hat \xi \approx .2\). This
means that the distribution of skyscraper heights has a heavy tail, and,
theoretically speaking, the maximum does not exist. While this is
obviously false---skyscrapers as defined in Section 1 are certainly
bounded above---it suggests sample averages and sums may be unreliable
for inference and extrapolation. Researchers must be careful
interpreting these quantities as evidence for their theories, especially
with small sample sizes. For example, the fact that the average height does not increase with
specific economic conditions may not indicate that the skyscraper
construction is unrelated to those conditions. Furthermore, increasing
average height in recent years may provide a poor basis for anticipating
heights in future years. Percentiles, such as the median, can be more stable
representatives of their theoretical analogs, and may prove better alternatives 
for conducting inference and extrapolating as demonstrated in the bottom-right
panel of Figure \ref{fig:qq}. Cirillo and Taleb (2016) have made similar points for
researchers using the total number of war casualties to determine whether
humans are less violent than in the past and whether wide-scale war will return
in the future.

Section 4 also found that the GPD parameters change little over time
after choosing an appropriate threshold. This suggests that the height
of the tallest skyscrapers is driven by the exponential increase in the
number of new buildings constructed each year and not a desire to build
the typical building taller. The distinction may be important for
researchers who use skyscraper heights as evidence of a cause-and-effect
relationship and then attempt to predict skyscraper heights from its cause. 
For example, our findings are consistent with the theory that the demand for 
extremely tall buildings led to the use of innovative technologies, such as 
faster elevators. Had the reverse been true---had the development of 
innovative technologies prompted extremely tall skyscrapers---an increase 
in skyscrapers would have been observed across the board as the technology
became available, and the GPD parameters would have changed substantially 
over time. While it is not the intent of this paper to draw any causal conclusions, 
we point out that researchers need to be careful of ``reverse-causality'',
attributing taller buildings as a consequence of a given factor instead
of its cause, without additional evidence. Spurious correlations
provide a poor basis for prediction.

Section 5 demonstrated how city-level effects might be estimated with a
hierarchical model. The dataset contains 258 cities, although the
typical city has only one skyscraper exceeding 150 meters and 40 floors.
Predictions are still possible for these cities because the model
borrows information across cities. Future researchers might benefit from
augmenting the hierarchy to include country and region effects. Or,
alternatively, covariate information, such as population and gross city
product at the time each skyscraper was completed, could be used
instead. These covariates would make the modeling assumptions more
plausible and may give insight into how cities might change policies
to increase or decrease skyscraper activity, provided covariates are chosen
judiciously and not based on spurious correlations. Spatiotemporal dependence
could even be modeled directly as discussed by Bao et al. (2006), Chan and
Gray (2006), and Ghil et al. (2011).

\subsection{7. Conclusion}\label{conclusion}

The major challenges confronting cities, such as sustainability, safety,
and equality, will depend on the infrastructure developed to accommodate
urbanization. Some urban planners have suggested that vertical
growth---the concentration of residents by constructing tall
buildings---be used to accommodate density. Others have argued that
urbanization will be too rapid to be accommodated by vertical growth
alone.

This paper finds that skyscraper construction will outpace urbanization
if present trends continue. Cities currently have around 800 skyscrapers
per billion people. By 2050, it is estimated that cities will have 6,800
skyscrapers per billion people. It also finds that the tallest among
these will be around fifty percent higher than those today and therefore
able to accommodate more people. However, these skyscrapers will not have
fifty percent more floors since the marginal capacity will diminish as
heights increase. For example, the one thousand meter building will have
seventy percent the floors of the mile-high building, despite being
sixty-two percent of the height.

This paper has not investigated whether skyscrapers will be constructed
in the cities with the most rapid urbanization. Nor has it investigated
whether skyscraper development should be used to accommodate density in
the first place. Instead, extreme value analysis provided a principled
basis to forecast future trends and quantify uncertainty. It
relies on the assumption that present trends continue, and there are a
variety of reasons why future trends may deviate from the past
sixty-eight years. For example, unprecedented technological changes may
result in new materials or methods that substantially reduce the cost of
construction. Other technological changes could cause a cultural shift
in how residents live or work, perhaps freeing up commercial space for
residential purposes. There is also the possibility of a hiatus due to
global upheaval, not unlike the period spanning the Great Depression and
Second World War. That period was ignored in this analysis, but future
work may choose to investigate this frightening prospect.

We conclude by stressing that extreme value analysis is one of many
principled strategies that could be used to predict skyscraper
development and the effects of urbanization more broadly. The previous
sections could be augmented by integrating theories from architecture,
engineering, policy, and social science. The incorporation of expert
knowledge is always useful, but it is particularly desirable with
extreme value analysis, as heavy tailed distributions are sensitive to
outliers and benefit from the context afforded by the theory
of other disciplines. More broadly, all disciplines will be necessary to
anticipate how cities will respond to the greatest migration in human
history, and solve perhaps the principal challenge of our time.

\subsection{8. Appendix}\label{appendix}

All models in Sections 4 and 5 are written in the Stan probabilistic programming
language (Carpenter et al. (2017)), and maximum likelihood estimates are
computed using the LBFGS algorithm in the R Core Team (2018) package
\texttt{rstan} (Stan Development Team (2018)). Standard errors are
computed from the observed Fisher information. Constrained parameters
such as scales and shapes are transformed so that they are unconstrained, the
Fisher information is calculated, and the constrained standard error is computed
using the delta method. Vehtari (2017) provides an introduction to univariate
Extreme Value Analysis with Stan. Section 2.6 of Carpenter et al. 
discusses the LBFGS algorithm for maximum likelihood estimation.

A 225 meter threshold for height (59 for floors) is utilized when fitting a GPD model.
Note that in the limit, \(\mu\) should be equal to the threshold \(u\). We allow 
\(\mu\) to vary (\(0 < \mu < \min(x_i)\)) in order to add flexibility to the model. This 
choice does not impact the conclusions in Section 4. Our point estimates match 
the output from the univariate \texttt{gpdFit} function in the package \texttt{fExtremes} 
(Wuertz, Setz, and Chalabi (2017)), which sets \(\mu = u\).

Before fitting the bivariate models, the number of floors was scaled by
3.8. The median skyscraper rises roughly 3.8 meters per floor, and
scaling by this number aligns the marginal distributions.

\subsection*{9. References}\label{references}
\addcontentsline{toc}{subsection}{9. References}

\hypertarget{refs}{}
\hypertarget{ref-al2012}{}
Al-Kodmany, Khei. 2012. ``The Logic of Vertical Density: Tall Buildings
in the 21st Century City.'' \emph{International Journal of High Rise
Buildings} 1 (2).

\hypertarget{ref-angel2011making}{}
Angel, Shlomo, Jason Parent, Daniel L Civco, and Alejandro M Blei. 2011.
``Making Room for a Planet of Cities.'' Lincoln Institute of Land Policy
Cambridge, MA.

\hypertarget{ref-ascher2011heights}{}
Ascher, Kate, and Rob Vroman. 2011. \emph{The Heights: Anatomy of a
Skyscraper}. Penguin Press London, England.

\hypertarget{ref-balkema1974residual}{}
Balkema, August A, and Laurens De Haan. 1974. ``Residual Life Time at
Great Age.'' \emph{The Annals of Probability}. JSTOR, 792--804.

\hypertarget{ref-barr2012skyscraper}{}
Barr, Jason. 2012. ``Skyscraper Height.'' \emph{The Journal of Real
Estate Finance and Economics} 45 (3). Springer: 723--53.

\hypertarget{ref-barr2017asia}{}
---------. 2017. ``Asia Dreams in Skyscrapers.'' \emph{The New York
Times}.

\hypertarget{ref-barr2016skyscrapers}{}
Barr, Jason, and Jingshu Luo. 2017. ``Economic Drivers: Skyscrapers in
Chinca.'' \emph{CTBUH Research Report}.

\hypertarget{ref-barr2015skyscraper}{}
Barr, Jason, Bruce Mizrach, and Kusum Mundra. 2015. ``Skyscraper Height
and the Business Cycle: Separating Myth from Reality.'' \emph{Applied
Economics} 47 (2). Taylor \& Francis: 148--60.

\hypertarget{ref-bao2006evaluating}{}
Bao, Yong, Tae-Hwy Lee, and Burak Saltoglu. 2006. ``Evaluating predictive performance
of value-at-risk models in emerging markets: a reality check''
\emph{Journal of forecasting} 25 (2). Wiley Online Library: 101--128.

\hypertarget{ref-canepari2014essay}{}
Canepari, Zack. 2014. ``Essay: A Planet of Suburbs.'' \emph{Economist}.

\hypertarget{ref-carpenter2017stan}{}
Carpenter, Bob, Andrew Gelman, Matthew D Hoffman, Daniel Lee, Ben
Goodrich, Michael Betancourt, Marcus Brubaker, Jiqiang Guo, Peter Li,
and Allen Riddell. 2017. ``Stan: A Probabilistic Programming Language.''
\emph{Journal of Statistical Software} 76 (1). Columbia Univ., New York,
NY (United States); Harvard Univ., Cambridge, MA (United States).

\hypertarget{ref-chan2006using}{}
Chan, Kam Fong and Philip Gray. 2006. ``Using extreme value theory to
measure value-at-risk for daily electricity spot prices' 
\emph{International Journal of Forecasting} 22 (2).
Elsevier: 283--300.

\hypertarget{ref-cirillo2016}{}
Cirillo, Pasquale and Nassim Nicholas Taleb. 2016. ``On the statistical properties 
and tail risk of violent conflicts'' \emph{Physica A: Statistical Mechanics and its 
Applications}. 452: 29--45.

\hypertarget{ref-clark1930skyscraper}{}
Clark, William Clifford, and John Lyndhurst Kingston. 1930. \emph{The
Skyscraper: Study in the Economic Height of Modern Office Buildings}.
American Institute of Steel.

\hypertarget{ref-cohen2006urbanization}{}
Cohen, Barney. 2006. ``Urbanization in Developing Countries: Current
Trends, Future Projections, and Key Challenges for Sustainability.''
\emph{Technology in Society} 28 (1-2). Elsevier: 63--80.

\hypertarget{ref-coles2001introduction}{}
Coles, Stuart, Joanna Bawa, Lesley Trenner, and Pat Dorazio. 2001.
\emph{An Introduction to Statistical Modeling of Extreme Values}. Vol.
208. Springer.

\hypertarget{ref-council2017skyscraper}{}
Council on Tall Buildings and Urban Habitat. 2017. ``The Skyscraper
Center.''

\hypertarget{ref-curl2015oxford}{}
Curl, James Stevens, and Susan Wilson. 2015. \emph{The Oxford Dictionary
of Architecture}. Oxford University Press, USA.

\hypertarget{ref-d2015wind}{}
D'Amico, Guglielmo, Filippo Petroni, and Flavio Prattico. 2015. ``Wind speed prediction for
wind farm applications by extreme value theory and copulas'' 
\emph{Journal of Wind Engineering and Industrial Aerodynamics} 145.
Elsevier: 229--236.

\hypertarget{ref-garreaud2018record}{}
Garreaud, RD. 2004. ``Record-breaking climate
anomalies lead to severe drought and environmental disruption in western Patagonia in 2016''
\emph{Climate Research} 74 (3): 217-229.

\hypertarget{ref-gencay2004extreme}{}
Gencay, Ramazan and Faruk Selcuk. 2004. ``Extreme 
value theory and Value-at-Risk: Relative performance in emerging markets'' 
\emph{International Journal of Forecasting} 20 (2).
Elsevier: 287--303.

\hypertarget{ref-ghil2011extreme}{}
Ghil M, P Yiou, S Hallegatte, BD Malamud, P Naveau, A Soloviev, P Friederichs, 
V Keilis-Borok, D Kondrashov, V Kossobokov, and O Mestre. 2011. ``Extreme 
events: dynamics, statistics and prediction''  \emph{Nonlinear Processes in Geophysics} 18 (3).
Copernicus: 295--350.

\hypertarget{ref-glaeser2011skyscrapers}{}
Glaeser, Edward. 2011. ``How Skyscrapers Can Save the City.'' \emph{The
Atlantic}.

\hypertarget{ref-gottmann1966skyscraper}{}
Gottmann, Jean. 1966. ``'Why the Skyscraper?'' \emph{Geographical
Review}. JSTOR, 190--212.

\hypertarget{ref-herrera2014modeling}{}
Herrera, Rodrigo and Nicol{\'a}s Gonz{\'a}lez. 2014. ``The modeling
and forecasting of extreme events in electricity spot markets'' 
\emph{International Journal of Forecasting} 30 (3).
Elsevier: 477--490.

\hypertarget{ref-huser2016likelihood}{}
Huser, Raphaël, Anthony C Davison, and Marc G Genton. 2016. ``Likelihood
Estimators for Multivariate Extremes.'' \emph{Extremes} 19 (1).
Springer: 79--103.

\hypertarget{ref-james2001architecture}{}
James, Steele. 2001. ``Architecture Today.'' \emph{Editura Phaidon, New
York}.

\hypertarget{ref-kashef2008race}{}
Kashef, Mohamad. 2008. ``The Race for the Sky: Unbuilt Skyscrapers.''
\emph{CTBUH Journal}, no. 1: 9--15.

\hypertarget{ref-Rquantreg2018}{}
Koenker, Roger. 2018. \emph{Quantreg: Quantile Regression}.
\url{https://CRAN.R-project.org/package=quantreg}.

\hypertarget{ref-lepik2004skyscrapers}{}
Lepik, Andres. 2004. \emph{Skyscrapers}. Prestel New York.

\hypertarget{ref-pickands1975statistical}{}
Pickands III, James. 1975. ``Statistical Inference Using Extreme Order
Statistics.'' \emph{The Annals of Statistics}. JSTOR, 119--31.

\hypertarget{ref-R2018}{}
R Core Team. 2018. \emph{R: A Language and Environment for Statistical
Computing}. Vienna, Austria: R Foundation for Statistical Computing.
\url{https://www.R-project.org/}.

\hypertarget{ref-rose2016well}{}
Rose, Jonathan FP. 2016. ``The Well-Tempered City: What Modern Science,
Ancient Civilizations, and Human Nature Teach Us About the Future of
Urban Life.'' Georgia Institute of Technology.

\hypertarget{ref-RStan2018}{}
Stan Development Team. 2018. ``RStan: The R Interface to Stan.''
\url{http://mc-stan.org/}.

\hypertarget{ref-sennott2004encyclopedia}{}
Sennott, R Stephen. 2004. \emph{Encyclopedia of 20th-Century
Architecture}. Routledge.

\hypertarget{ref-swilling2016curse}{}
Swilling, Mark. 2016. ``The Curse of Urban Sprawl: How Cities Grow, and
Why This Has to Change.'' \emph{The Guardian}.

\hypertarget{ref-tawn1988bivariate}{}
Tawn, Jonathan A. 1988. ``Bivariate Extreme Value Theory: Models and
Estimation.'' \emph{Biometrika} 75 (3). Oxford University Press:
397--415.

\hypertarget{ref-thompson2017high}{}
Thompson, Vikki, Nick J Dunstone, Adam A Scaife,Doug M Smith, 
Julia M Slingo, Simon Brown, and Stephen E Belcher 2017. ``High risk of 
unprecedented UK rainfall in the current climate''  \emph{Nature communications} 8 (1).
Nature Publishing Group: 107.

\hypertarget{ref-un2015world}{}
UN, DESA. 2015. ``World Urbanization Prospects: The 2014 Revision.''
\emph{United Nations Department of Economics and Social Affairs,
Population Division: New York, NY, USA}.

\hypertarget{ref-un2018world}{}
---------. 2018. ``World Urbanization Prospects: The 2018 Revision, Key
Facts.'' \emph{United Nations Department of Economics and Social
Affairs, Population Division: New York, NY, USA}.

\hypertarget{ref-vehtari2017gpd}{}
Vehtari, Aki. 2017. ``Extreme Value Analysis and User Defined
Probability Functions in Stan.''

\hypertarget{ref-willis1995form}{}
Willis, Carol. 1995. \emph{Form Follows Finance: Skyscrapers and
Skylines in New York and Chicago}. Princeton Architectural Press.

\hypertarget{ref-RfExtremes2018}{}
Wuertz, Diethelm, Tobias Setz, and Yohan Chalabi. 2017. \emph{FExtremes:
Rmetrics - Modelling Extreme Events in Finance}.
\url{https://CRAN.R-project.org/package=fExtremes}.

\end{document}